\documentstyle[aps,epsf]{revtex}

\newcommand{\be}{\begin{equation}}
\newcommand{\ee}{\end{equation}}
\def\la{\mathrel{\mathpalette\fun <}}
\def\ga{\mathrel{\mathpalette\fun >}}
\def\fun#1#2{\lower3.6pt\vbox{\baselineskip0pt\lineskip.9pt
\ialign{$\mathsurround=0pt#1\hfil##\hfil$\crcr#2\crcr\sim\crcr}}}
\begin{document}
\draft
\title{Proton spin content and QCD topological susceptibility}
\author{B. L. Ioffe and A.G.Oganesian}
\address{Institute of Theoretical
and Experimental Physics, B.Cheremushkinskaya 25,
Moscow 117218, Russia}
\date{\today}
\maketitle
\begin{abstract}
The part of the proton spin $\Sigma$ carried by $u, d, s$ quarks is
calculated in the framework of the QCD sum rules in the external
fields. The operators up to dimension 9 are accounted.  An important
contribution comes from the operator of dimension 3, which in the limit
of massless $u, d, s$ quarks is equal to the derivative
of QCD topological  susceptibility $\chi^{\prime} (0)$. The comparison
with the experimental data on $\Sigma$ gives $\chi^{\prime}(0)= (2.3
\pm 0.6) \times 10^{-3} ~ GeV^2$. The limits on $\Sigma$ and
$\chi^{\prime}(0)$ are found from selfconsistency of the sum rule,
$\Sigma \ga 0.05,~~ \chi^{\prime} (0) \ga 1.6 \times 10^{-3} ~ GeV^2$. The
values of $g_A = 1.37 \pm 0.10$ and $g^8_A = 0.65 \pm 0.15$ are also
determined.

\end{abstract}

\widetext

In the last years, the problem of nucleon spin content and particularly
the question which part of the nucleon spin is carried by quarks,
attracts a strong interest. The valuable information comes from the
measurements of the spin-dependent nucleon structure functions $g_1(x,
Q^2)$ in deep inelastic $e(\mu)N$ scattering (for the recent data see
[1,2], for a review \cite{3}). The parts of the nucleon spin carried by
$u, d$
and $s$-quarks are determined from the measurements of the first moment
of $g_1(x, Q^2)$

\begin{equation}
\Gamma_{p,n}(Q^2) = \int \limits_{0}^{1} dx g_{1;p,n} (x, Q^2)
\end{equation}
The data allows one to find the value of $\Sigma$ -- the part of
nucleon spin carried by three flavours of light quarks
$\Sigma = \Delta u + \Delta d + \Delta s$,
where $\Delta u, \Delta d, \Delta s$ are the parts of nucleon spin
carried by $u,d,s$ quarks. On the basis of the operator product expansion
(OPE) $\Sigma$ is related to the proton matrix element of the flavour
singlet axial current $j^0_{\mu 5}$

\begin{equation}
2 ms_{\mu} \Sigma = \langle p, s \vert j^0_{\mu 5} \vert p,s \rangle,
\end{equation}
where $s_{\mu}$ is the proton spin 4-vector, $m$ is the proton mass.
The renormalization scheme in the calculation of perturbative QCD
corrections to $\Gamma_{p,n}$ can be arranged in such a way that
$\Sigma$ is scale independent.

An attempt to calculate $\Sigma$ using QCD sum rules in external fields
was done in ref.\cite{4}. Let us shortly recall the idea. The polarization operator

\begin{equation}
\Pi(p) = i \int d^4 x e^{ipx} \langle 0 \vert T \{ \eta(x),
\bar{\eta} (0) \} \vert 0 \rangle
\end{equation}
was considered, where
\begin{equation}
\eta(x) = \varepsilon^{abc} \Biggl (u^a(x) C \gamma_{\mu} u ^b(x)
\Biggr ) \gamma _{\mu} \gamma_5 d^c(x)
\end{equation}
is the current with proton quantum numbers [5], $u^a, d^b$ are quark
fields, $a,b,c$ are colour indeces. It is assumed that the term
\begin{equation}
\Delta L = j^0_{\mu 5} A_{\mu}
\end{equation}
where $A_{\mu}$ is a constant singlet axial field, is added to QCD Lagrangian.
In the weak axial field approximation $\Pi(p)$ has the form

\begin{equation}
\Pi(p) = \Pi^{(0)} (p) + \Pi^{(1)}_{\mu} (p) A_{\mu}.
\end{equation}
$\Pi^{(1)}_{\mu}(p)$ is calculated in QCD by
OPE at $p^2 < 0, \vert
p^2 \vert \gg R^{-2}_c$, where $R_c$ is the confinement radius. On the
other hand,  using dispersion relation, $\Pi^{(1)}_{\mu} (p)$ is
represented by the contribution of the physical states, the lowest of
which is the proton state. The contribution of excited states is
approximated as a continuum and suppressed by the Borel transformation. The
desired answer is obtained by equalling of these two representations. This
procedure can be applied to any Lorenz structure of $\Pi^{(1)}_{\mu} (p)$ ,
but as was argued in [6,7], the best accuracy can be obtained by considering
the chirality conserving structure $2 p_{\mu} \hat{p} \gamma_5$ .

An essential ingredient of the method is the appearance of induced by
the external field vacuum expectation values (v.e.v). The most
important of them in the problem at hand is

\begin{equation}
\langle 0 \vert j^0_{\mu 5} \vert 0 \rangle_0 \equiv 3 f^2_0 A_{\mu}
\end{equation}
of dimension 3. The constant $f^2_0$ is related to QCD topological
susceptibility. Using (5), we can write
$$
\langle 0 \vert j^0_{\mu 5} \vert 0 \rangle_A = lim_{q \to 0} ~ i \int~
d^4 xe^{iqx} \langle 0 \vert T \{ j^0_{\nu 5} (x), j^0_{\mu 5} (0) \}
\vert 0 \rangle A_{\nu} \equiv
$$
\begin{equation}
\equiv lim_{q \to 0} P_{\mu \nu} (q) A_{\nu}
\end{equation}
The general structure of $P_{\mu \nu} (q)$ is

\begin{equation}
P_{\mu \nu} (q) = -P_L(q^2) \delta_{\mu \nu} + P_T(q^2) (-\delta_{\mu
\nu} q^2 + q_{\mu} q_{\nu})
\end{equation}
Because of anomaly there are no massless states in the spectrum of the
singlet polarization operator $P_{\mu \nu}$ even for massless quarks.
$P_{T,L}(q^2)$ also have no kinematical singularities at $q^2 = 0$ .
Therefore, the nonvanishing value $P_{\mu \nu} (0)$ comes entirely from
$P_L(q^2)$. Multiplying $P_{\mu \nu} (q)$ by $q_{\mu} q_{\nu}$, in the
limit of massless $u, d, s$ quarks we get

$$
q_{\mu} q_{\nu} P_{\mu \nu} (q) = -P_L (q^2) q^2 = N^2_f
(\alpha_s / 4 \pi)^2 i \int~ d^4 xe^{iqx} \times
$$
\begin{equation}
\times \langle 0 \vert T {G^n_{\mu \nu} (x) \tilde{G}^n_{\mu \nu} (x),
G^m_{\lambda \sigma} (0) \tilde{G}^m _{\lambda \sigma} (0)} \vert 0 \rangle,
\end{equation}
where $G^n_{\mu \nu}$ is the gluonic field strength, $\tilde{G}_{\mu
\nu} = (1/2) \varepsilon_{\mu \nu \lambda \sigma} G_{\lambda \sigma}$.(The anomaly
condition was used, $N_f = 3$.).
Going to the limit $q^2 \to 0$, we have

\begin{equation}
f^2_0 = -(1/3) P_L(0) = \frac{4}{3} N^2_f \chi^{\prime} (0),
\end{equation}
where $\chi(q^2)$ is the topological susceptibility

\begin{equation}
\chi(q^2) = i~ \int d^4 x e^{iqx} \langle 0 \vert T {Q_5 (x),
Q_5 (0)} \vert 0 \rangle
\end{equation}

\begin{equation}
Q_5(x) = (\alpha_s / 8 \pi)~ G^n_{\mu \nu} (x) \tilde{G}^n _{\mu
\nu} (0),
\end{equation}
As is well known (see, e.g., the review \cite{8}), $\chi(0) = 0$ if there is
at least one massless quark. The attempt to find $\chi^{\prime}(0)$ itself
by QCD sum rules failed: it was found \cite{4} that OPE does not converge
in the domain of characteristic scales for this problem. However, it
was possible to derive the sum rule, expressing $\Sigma$ in terms of
$f^2_0$ (7) or $\chi^{\prime}(0)$. The OPE up to dimension $d=7$ was
performed in ref.\cite{4}. Among the induced by the external field v.e.v.'s
besides (7), the v.e.v. of the dimension 5 operator

\begin{equation}
g \langle 0 \vert \sum \limits_{q} \bar{q} \gamma_{\alpha} (1/2) \lambda
^n \tilde{G}^n_{\alpha \beta} q \vert 0 \rangle _A \equiv 3 h_0
A_{\beta}, ~~~ q = u, d, s
\end{equation}
was accounted and the constant $h_0$ was estimated using a special sum
rule,\\  $h_0 \approx 3 \times 10^{-4} GeV^4$ . There were also accounted the
gluonic condensate $d = 4$ and the square of quark condensate $d= 6$
(both times the external $A_{\mu}$ field operator, $d=1$). However, the
accuracy of the calculation was not good enough for reliable
calculation of $\Sigma$ in terms of $f^2_0$: the necessary requirement
of the method -- the weak dependence of the result on the Borel
parameter was not well satisfied.

In this paper we improve the accuracy of the calculation by going to
higher order terms in OPE up to dimension 9 operators. Under the
assumption of factorization -- the saturation of the product of
four-quark operators by the contribution of an intermediate vacuum
state -- the dimension 8 v.e.v.'s are accounted (times $A_{\mu}$):

\begin{equation}
-g \langle 0 \vert \bar{q} \sigma_{\alpha \beta} (1/2) \lambda^n
G^n_{\alpha \beta} q \cdot \bar{q} q \vert 0 \rangle = m^2_0 \langle 0
\vert \bar{q} q \vert 0 \rangle^2,
\end{equation}
where $m^2_0 = 0.8 \pm 0.2 ~GeV^2$
was determined in \cite{9}.
In the framework of the same factorization hypothesis the induced by
the external field v.e.v. of dimension 9

\begin{equation}
\alpha_s \langle 0 \vert j^{(0)}_{\mu 5} \vert 0 \rangle_A \langle 0
\vert \bar{q} q \vert 0 \rangle^2
\end{equation}
is also accounted. In the calculation we used the following expression for
the quark Green function in the constant external axial field  \cite{7}:
$$
\langle 0 \vert T \{ q^a_{\alpha}(x),~ \bar{q}^b_{\beta} (0) \} \vert 0
\rangle_A = i \delta^{ab} \hat{x}_{\alpha \beta} / 2 \pi^2 x^4 +
$$
$$
+ (1/2 \pi^2) \delta^{ab} (A x) (\gamma_5 \hat{x})_{\alpha \beta} / x^4
- (1/12) \delta^{ab} \delta_{\alpha \beta} \langle 0 \vert \bar{q} q
\vert 0 \rangle +
$$
$$
+(1/72)i\delta^{ab} \langle 0 \vert \bar{q} q \vert 0 \rangle (\hat{x}
\hat{A} \gamma_5 - \hat{A} \hat{x} \gamma_5)_{\alpha \beta} +
$$
\begin{equation}
+ (1/12) f^2_0 \delta^{ab} (\hat{A} \gamma_5)_{\alpha \beta} + (1/216)
  \delta^{ab} h_0 \Biggl [(5/2) x^2 \hat{A} \gamma_5 - (A x) \hat{x}
\gamma_5 \Biggr ]_{\alpha \beta}
\end{equation}
The terms of the third power  in $x$-expansion of
quark propagator proportional to $A_{\mu}$ are omitted in (17), because they do not contribute to the
 tensor structure of $\Pi_{\mu}$ of interest. Quarks are considered to be
 in the constant external gluonic field and quark and gluon QCD equations
 of motion are exploited (the related formulae are given in [10]). There is also an
 another source of v.e.v. $h_0$  to appear besides the $x$-expansion
 of quark propagator given in eq.(17): the quarks in the condensate absorb
 the soft gluonic field emitted by
 other quark. A similar
 situation takes place also in the calculation of the v.e.v. (16)
 contribution. The accounted diagrams with dimension 9 operators have no
 loop integrations.  There are others v.e.v.  of dimensions $d \leq 9$
 particularly containing gluonic fields.  All of them, however, correspond
 to at least one loop integration and are suppressed by the numerical factor
 $(2 \pi)^{-2}$. For this reason they are disregarded.

The sum rule for $\Sigma$ is given by
$$
\Sigma + C_0 M^2 = -1 + \frac{8}{9 \tilde{\lambda}^2_N} e^{m^2/M^2}
\left \{a^2 L^{4/9} + \right.
$$
\begin{equation}
\left. + 6\pi^2 f^2_0 M^4 E_1 \Biggl (\frac{W^2}{M^2} \Biggr ) L^{-4/9} +
14 \pi^2 h_0 M^2 E_0 \Biggl (\frac {W^2}{M^2} \Biggr ) L^{-8/9} -
\frac{1}{4}~ \frac{a^2 m^2_0}{M^2} - \frac{1}{9} \pi \alpha_s f^2_0~
\frac{a^2}{M^2} \right \}
\end{equation}

Here $M^2$ is the Borel parameter, $\tilde{\lambda}_N$ is defined as
$\tilde{\lambda}^2_N = 32 \pi^4 \lambda^2_N = 2.1 ~ GeV^6$,
$\langle 0 \vert \eta \vert p \rangle = \lambda_N v_p,$
where $v_p$ is proton spinor, $W^2$ is the continuum threshold, $W^2 =
2.5 ~GeV^2$,

\begin{equation}
a = -(2 \pi)^2 \langle 0 \vert \bar{q} q \vert 0 \rangle = 0.55 ~ GeV^3
\end{equation}
$$
E_0(x) = 1 - e^{-x} , ~~~ E_1(x) = 1 - (1 + x)e^{-x}
$$
$L = ln (M/\Lambda)/ ln (\mu/\Lambda),~~~
\Lambda = \Lambda_{QCD} = 200 ~ MeV$ and the normalization point  $\mu$
was chosen $\mu = 1 ~ GeV$. When deriving (18) the sum rule for the
nucleon mass was exploited what results in  appearance of the first
term, -1, in the right hand side (rhs) of (18). This term absorbs the contributions of
the bare loop, gluonic condensate as well as $\alpha_s$ corrections
to them and essential part of terms, proportional to $a^2$ and $m^2_0 a^2$.
The values of the parameters, $a, \tilde{\lambda}^2_N, W^2$ taken
above were chosen by the best fit of the sum rules for the nucleon mass
(see [11], Appendix B) performed at $\Lambda = 200 ~ MeV$. It can be shown,
using the value of the ratio $2m_s/ (m_u + m_d) = 24.4 \pm 1.5$  [12] that
$a(1~ GeV) = 0.55 ~ GeV^3$ corresponds to $m_s(1 ~ GeV) = 153 ~ MeV$.
$\alpha_s$ corrections are accounted in the leading order (LO) what results
in appearance of anomalous dimensions. Therefore $\Lambda$ has the meaning
of effective $\Lambda$ in LO. The unknown constant $C_0$ in the left-hand
side (lhs) of (18) corresponds to the contribution of inelastic transitions
$p \to N^* \to interaction~ with A_{\mu} \to p$ (and in inverse order). It
cannot be determined theoretically and may be found from $M^2$ dependence of
the rhs of (18) (for details see  [11,13]). The necessary condition of the
validity of the sum rule is $\vert \Sigma \vert \gg \vert C_0 M^2 \vert exp
[(-W^2 + m^2)/M^2]$ at characteristic values of $M^2$  [13]. The
contribution of the last term in the rhs of (18) is negligible.
The sum rule (18) as well as the sum
rule for the nucleon mass is reliable in the interval of the Borel parameter
$M^2$  where the last term of OPE is small less than $10-15\%$ of the
total and the contribution of continuum does not exceed $40-50\%$ . This
fixes the interval $0.85 < M^2 < 1.4 ~ GeV^2$.The $M^2$-dependence of the
rhs of (18) at $f^2_0 = 3 \times 10^{-2}~ GeV^2$ is plotted in Fig.1. The
complicated expression in rhs of (18) is indeed an almost linear function of
$M^2$ in the given interval! This fact strongly supports the
reliability of the approach. The best values of $\Sigma =
\Sigma^{fit}$ and $C_0 = C^{fit}_0$ are found from the $\chi^2$
fitting procedure

\begin{equation}
\chi^2 = \frac{1}{n} \sum \limits ^{n}_{i=1}~ [\Sigma^{fit} -
C^{fit}_0 M^2_i - R(M^2_i)]^2 = min,
\end{equation}
where $R(M^2)$  is the rhs of (18).

The values of $\Sigma$ as a function of $f^2_0$  are
plotted in Fig.2 together with $\sqrt{\chi^2}$.
In our approach the gluonic
contribution cannot be separated and is included in $\Sigma$. The
experimental value of $\Sigma$ can be estimated [1,2] (for discussion
see \cite{14}) as $\Sigma = 0.3 \pm 0.1$. Then from Fig.2 we have $f^2_0 =
(2.8 \pm 0.7) \times 10^{-2} ~ GeV^2$ and $\chi^{\prime}(0) = (2.3 \pm 0.6)
 \times 10^{-3}~  GeV^2$ . The error in $f^2_0$ and $\chi^{\prime}$ besides
the experimentall error includes the uncertainty in the sum rule estimated
as equal to the contribution of the last term in OPE (two last terms in
Eq.18)
and a possible role of NLO $\alpha_s$ corrections.
At $f^2_0 < 0.02~ GeV^2$  $\chi^2$ is much worse and the fit becomes
unstable.  This allows us to claim (with some care, however,) that
$\chi^{\prime} (0) \geq 1.6 \times 10^{-3} GeV^2$ and $\Sigma \geq 0.05$ from
the requirement of selfconsistency of the sum rule. The $\chi^2$ curve also
favours an upper limit for $\Sigma \la 0.6$.  At $f^2_0 = 2.8 \times
10^{-2}~ GeV^2$  the value of the constant $C_0$ found from the fit is $C_0
= 0.19 ~ GeV^{-2}$.  Therefore, the mentioned above necessary condition of
the sum rule validity is well satisfied. Recently, the first attempt to
calculate $\chi^{\prime}(0)$ on the lattice was performed \cite{15}. The
result is $\chi^{\prime}(0) = (0.4 \pm 0.2) \times 10^{-3}~  GeV^2$, much
below our value.  However, as mentioned by the authors, the calculation has
some drawbacks and the result is preliminary.

From the same sum rule (18) it is possible to find $g^8_A$ -- the
proton coupling constant  with the octet axial current, which enters the QCD
formula for $\Gamma_{p,n}$ \cite{3}. There are two differences in
comparison with (18):

I. Instead of $f^2_0$ it appears the square $f^2_8$
of the pseudoscalar meson coupling constant with the octet axial
current. In the limit of strict SU(3) flavour symmetry it is equal to
$f^2_{\pi}$,
$f_{\pi} = 133 ~ MeV$. However, it
is known, that SU(3) symmetry is violated and the kaon
decay constant, $f_K \approx 1.25 f_{\pi}$ . In the linear in $s$-quark
mass $m_s$ approximation $f_{\eta} = 1.31 f_{\pi}$.
We put for $f^2_8$ the value $f^2_8 = 2.6 \times 10^{-2}~
GeV^2$, intermediate between $f^2_{\pi}$ and $f^2_{\eta}$ .

2. $h_0$ should be substituted by $m^2_1 f^2_{\pi}$. The constant
$m^2_1$ is determined by the sum rules suggested in \cite{16}. A new fit
corresponding to the values of the parameters used above, was
performed and it was found; $m^2_1 = 0.16~  GeV^2$.

The $M^2$ -dependence of $g^8_A + C_8 M^2$ is presented in Fig.1 and
the best fit according to the fitting procedure (20) at $1.0 \leq M^2
\leq 1.3 ~ GeV^2$ gives

\begin{equation}
g^8_A = 0.65 \pm 0.15, ~~~ C_8 = 0.10 ~GeV^{-2} ~~~\sqrt{\chi^2} = 1.2
\times 10^{-3}
\end{equation}
(The error includes the uncertainties in the
sum rule as well as in the value of $f^2_8$). The obtained value of $g^8_A$
within the errors coincides with $g^8_A = 0.59 \pm 0.02$ \cite{17} found
from the data on baryon octet $\beta$-decays under assumption of strict
SU(3) flavour symmetry and contradicts the hypothesis of bad violation of
SU(3) symmetry in baryon axial octet coupling constants \cite{18}.

A similar sum rule with the account of dimension 9 operators can be
derived also for $g_A$ -- the nucleon axial $\beta$-decay coupling
constant. It is an extension of the sum rule found in [6] and has the
form

\begin{equation}
g_A + C_A M^2 = 1 + \frac{8}{9 \tilde{\lambda}^2_N} e^{m^2/M^2} \Biggl
[a^2 L^{4/9} + 2 \pi^2 m^2_1 f^2_{\pi} M^2 - \frac{1}{4} a^2
\frac{m^2_0}{M^2} + \frac{5}{3} \pi \alpha_s f^2_{\pi} \frac{a^2}{M^2} \Biggr ]
\end{equation}
The main term in OPE of dimension 3 proportional to $f^2_{\pi}$
occasionally was cancelled. For this reason the higher order
terms of OPE may be more important in the sum rule for $g_A$ than in the previous
ones. The $M^2$ dependence of $g_A - 1 + C_A M^2$ is plotted in Fig.1,
lower curve; the curve is almost the straight line, as it should be.
The best fit gives

\begin{equation}
g_A = 1.37 \pm 0.10, ~~~ C_A = -0.088~ GeV^{-2}, ~~~~ \sqrt{\chi^2} =
1.0 \times 10^{-3}
\end{equation}
in comparison with the world average $g_A
= 1.260 \pm 0.002$ [19]. The inclusion of dimension 9 operator contribution
essentially improves the result: without it $g_A$ would be about 1.5 and
$\chi^2$ would be much worse.

\acknowledgments

We are thankful to H.Leutwyler for useful discussions and for the 
hispitality at the Bern University. The work was supported in part by 
CRDF Grant RP2-132, INTAS Grant 93-0283, RFFR Grant 97-02-16131 and 
Swiss Grant 7SUPJ048716.


\begin{figure}
\centerline{\epsfxsize=0.7\textwidth\epsfbox{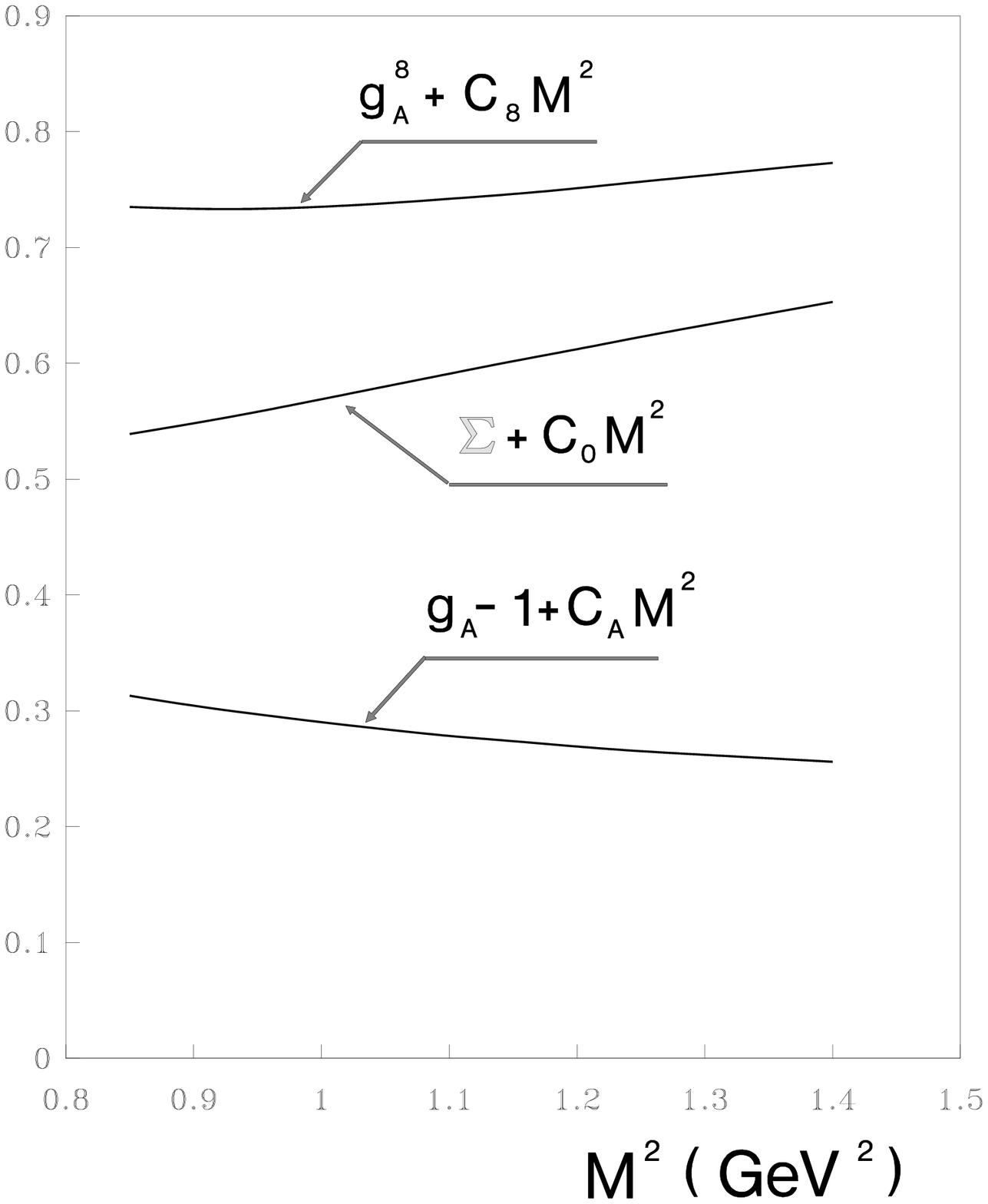}}
\caption{The $M^2$-dependence of $\Sigma + C_0 M^2$ at $f^2_0 = 3 
\times 10^{-2}~ GeV^2$ , eq.18, $g^8_A + C_8 M^2$, and $g_A - 1 + C_A 
M^2$, eq.22.}
\end{figure}

\begin{figure}
\centerline{\epsfxsize=0.7\textwidth\epsfbox{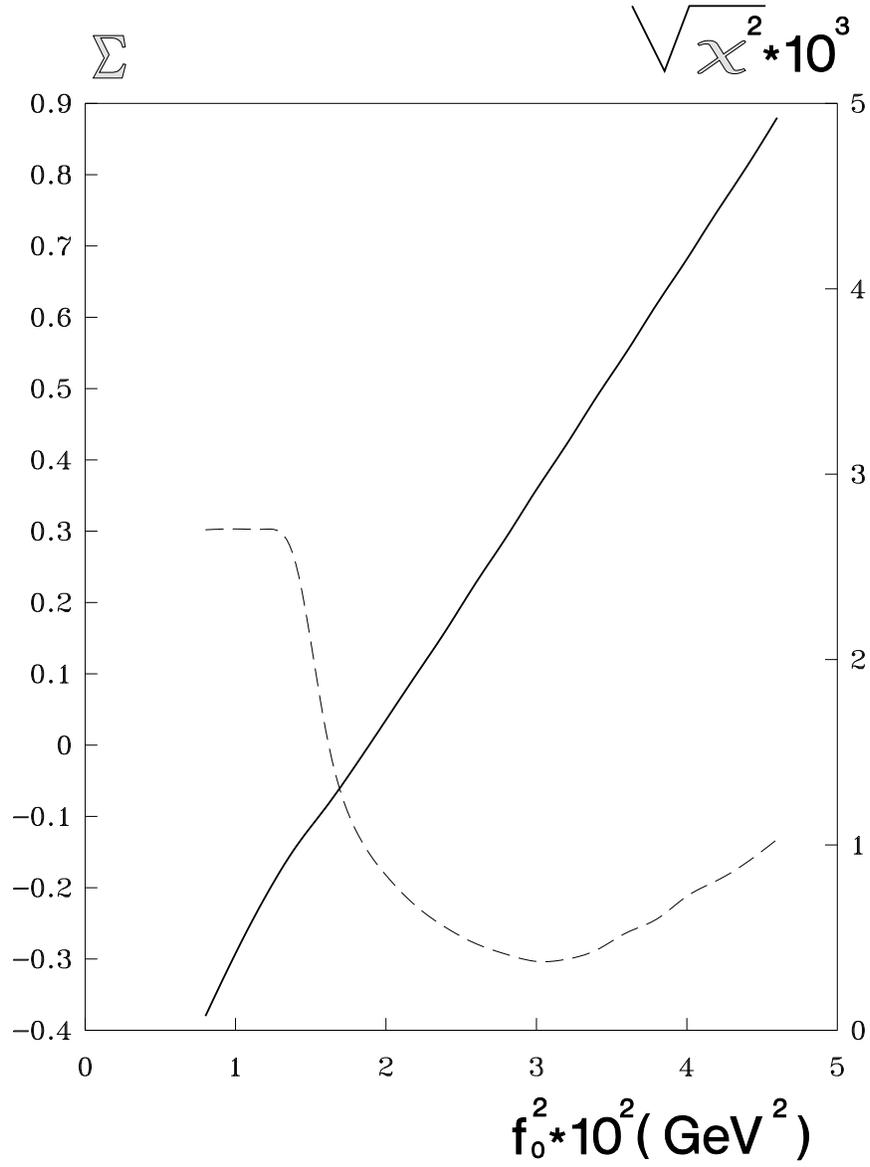}}
\caption{$\Sigma$  (solid line, left ordinate
axis) and $\sqrt{\chi^2}$, eq.(20), (dashed line, right ordinate 
axis) as a functions of $f^2_0$.}
\end{figure}

\end{document}